\documentclass[12pt]{article}
\usepackage{amsmath,amsthm,amssymb,euscript,epsfig}
\usepackage{array}
\usepackage{fancybox}
\usepackage[usenames]{color}
\usepackage{amsfonts,bm}
\usepackage{graphicx}

\newcommand{\be}{\begin{equation}}
\newcommand{\ee}{\end{equation}}
\newcommand{\bea}{\begin{eqnarray}\displaystyle}
\newcommand{\eea}{\end{eqnarray}}

\makeatletter
\@addtoreset{equation}{section}
\makeatother
\renewcommand{\theequation}{\thesection.\arabic{equation}}
\def\one{{\hbox{ 1\kern-.8mm l}}}
\def\zero{{\hbox{ 0\kern-1.5mm 0}}}

\begin{document}
\makeatletter
\@addtoreset{equation}{section}
\makeatother
\renewcommand{\theequation}{\thesection.\arabic{equation}}

\rightline{WITS-CTP-085}
\vspace{1.8truecm}

\vspace{15pt}


{\LARGE
\centerline{\bf  Non-planar Anomalous Dimensions}
\centerline{\bf  in the $sl(2)$ Sector}
}  

\vskip.5cm 

\thispagestyle{empty} \centerline{
    {\large \bf Robert de Mello Koch\footnote{{\tt robert@neo.phys.wits.ac.za}},}
   {\large \bf Pablo Diaz\footnote{{\tt Pablo.DiazBenito@wits.ac.za}}
    and Hesam Soltanpanahi\footnote{{\tt Hesam.SoltanpanahiSarabi@wits.ac.za}}}
                                                       }

\vspace{.4cm}
\centerline{{\it National Institute for Theoretical Physics ,}}
\centerline{{\it Department of Physics and Centre for Theoretical Physics }}
\centerline{{\it University of Witwatersrand, Wits, 2050, } }
\centerline{{\it South Africa } }

\vspace{1.4truecm}

\thispagestyle{empty}

\centerline{\bf ABSTRACT}

\vskip.4cm 

In this note we compute the non-planar one loop anomalous dimension of restricted Schur polynomials that
belong to the $sl(2)$ sector of ${\cal N}=4$ super Yang-Mills theory and have a bare dimension of order $N$. 
Although the details are rather different, ultimately the problem of diagonalizing the dilatation operator 
in the $sl(2)$ sector can be reduced to the $su(2)$ sector problem. In this way we establish the expected 
dynamical emergence of the Gauss Law for giant gravitons and further show that the dilatation operator 
reduces to a set of decoupled harmonic oscillators.

\setcounter{page}{0}
\setcounter{tocdepth}{2}

\newpage

\setcounter{footnote}{0}

\linespread{1.1}
\parskip 4pt

\section{Introduction}

The most detailed and tested realization of the AdS/CFT correspondence postulates a complete equivalence between
${\cal N}=4$ super Yang-Mills theory with gauge group $U(N)$ and type IIB string theory on the AdS$_5\times$S$^5$
background with $N$ units of five form flux\cite{Maldacena:1997re}. The correspondence has motivated outstanding progress
in the computation of two point functions in ${\cal N}=4$ super Yang-Mills theory. In particular, integrable structures
governing the anomalous dimensions of the theory in the planar limit have been discovered\cite{Minahan:2002ve,Beisert:2010jr}. 
In much of the literature on the subject, the planar limit and large $N$ limit are usually taken to mean the same thing. This
is not true\cite{Balasubramanian:2001nh}.
An interesting question is whether or not integrability is present in other large $N$ ({\it but not planar}!) limits of the theory.
In this work we will have some relevant comments on this issue. 

Operators constructed using a single complex scalar field $Z$ transforming in the adjoint of $U(N)$ are ${1\over 2}$-BPS.
Homogeneous polynomial operators in $Z$ (we have the Schur polynomials in mind) have ${\cal R}$-charge given by the degree 
of the polynomial. It is possible to establish a dictionary between these operators and ${1\over 2}$-BPS objects in the dual 
string theory\cite{Corley:2001zk,Berenstein:2004kk,Lin:2004nb}. The entries in the dictionary are organized by ${\cal R}$-charge. 
Operators with an ${\cal R}$-charge of order $N$ are dual to giant gravitons\cite{McGreevy:2000cw}. Among these, Schur polynomials 
labeled by Young diagrams with $O(1)$ long\footnote{Long means order $N$ boxes.} columns are dual to sphere giant 
gravitons\cite{Balasubramanian:2002sa} while Schur polynomials labeled by Young diagrams with $O(1)$ long rows are dual to AdS 
giant gravitons\cite{Corley:2001zk}.

Excited giant graviton states, which are not ${1\over 2}$-BPS can be described in terms of open strings which end on the giant. 
Operators dual to excited giant 
gravitons were proposed in \cite{Balasubramanian:2004nb}. Since giant gravitons have a compact world volume, Gauss' Law forces the 
total charge on the worldvolume to vanish\cite{Sadri:2003mx}. A highly non-trivial test of the proposal of \cite{Balasubramanian:2004nb} 
is that the number of operators that can be defined matches the number of states obeying this Gauss Law constraint. 
The two point functions of these operators, the {\it restricted Schur polynomials}, were computed exactly, in the free field theory limit, 
in \cite{Bhattacharyya:2008rb}, by exploiting the technology developed in \cite{de Mello Koch:2007uu,de Mello Koch:2007uv,Bekker:2007ea}. 
It was also shown that the restricted Schur polynomials provide a basis for the gauge invariant
local operators built using only scalar (adjoint Higgs) fields\cite{Bhattacharyya:2008rc}. 
Numerical studies of the dilatation operator, when acting on decoupled sectors within the $su(2)$ sector of the theory, 
that have a sphere giant graviton number equal to two showed that the spectrum of the dilatation operator is that of a 
set of decoupled harmonic oscillators \cite{Koch:2010gp,VinceKate}. Analytic studies of the dilatation operator in these 
sectors within the $su(2)$ sector of the theory, with either two sphere giants or two AdS giants\cite{bhw}
and then in general\cite{dgm} have demonstrated that the spectrum of the dilatation operator is that of a 
set of decoupled harmonic oscillators and further, that the Gauss' Law emerges at one loop. This dynamical emergence
of the Gauss Law significantly extends the counting arguments of \cite{Balasubramanian:2004nb}.
A recent study has shown that the frequency of the decoupled oscillators emerging from the dilatation operator
can be obtained from a coupled system of springs which lends further support to the idea that these operators are
dual to an open string system\cite{gs}.

In this note we will study operators dual to excited AdS giants.  The strings carry an angular momentum along an $S^1\subset $AdS$_5$ 
direction. The gauge theory description of strings spinning along the AdS directions requires us to consider the action of covariant 
derivatives on the set of scalar operators. In this way we will be able to investigate whether or not the dilatation operator reduces
to a set of decoupled oscillators as well as if the Gauss Law emerges dynamically in the $sl(2)$ sector of the theory. It will be enough to
consider operators built using a single complex scalar field $Z$ and covariant derivatives taken in a unique 
spacetime direction, acting on it. The basic building blocks $D_+^p Z$ transform in the infinite dimensional
spin $j=-{1\over 2}$ representation of a non-compact $sl(2)$ subsector of the conformal subalgebra. In the 
next section we will describe how to build a restricted Schur polynomial basis for a $sl(2)$ sector. We will then
use the one-loop dilatation operator of ${\cal N}=4$ super Yang-Mills theory constructed by Beisert\cite{beisert} 
and restrict it to the relevant $sl(2)$ sector. The result, given in section 3, is the main result of this note. 
Section 4 is reserved for a discussion of our results. In order to control the length of this article, we have not
attempted to make it completely self contained. We refer the reader to \cite{de Mello Koch:2007uu,dgm} for background.

\section{Restricted Schur Polynomials for the $sl(2)$ Sector}

This section will describe a restricted Schur polynomial basis for operators built using 
$n_z$ $Z$'s and $m$ vector ``impurities'', that is, $m$ covariant derivatives $D_+$ act 
on the $n_Z$ $Z$ fields. These operators do not mix with other operators under the action of the 
dilatation operator - they form the closed $sl(2)$ subsector\cite{beisert}. We will build our operators 
out of $Z_{(i)}$, $i=0,1,2,...,m$ where\footnote{This inner product is defined with respect 
to the Zamolodchikov metric (defined in equation (\ref{ZamMet})) on the space of local operators
of the conformal field theory.} 
$$
  Z^{(n)}={1\over n!}D_+^n Z,\qquad Z^{(n)\,\,\dagger} = {1\over n!}D_-^n Z^\dagger
$$
and $Z^{(0)}\equiv Z$. The action of the covariant derivative is completely specified by
$$
 D_+^n Z = D_+ Z^{(n-1)}=\partial_+ Z^{(n-1)} -ig_{YM}\big[ A_+,Z^{(n-1)}\big]\, .
$$
Denote the number of $Z^{(i)}$ in the operator by $n_i$. We have
$$
  \sum_{i=0}^m i n_i =m\, ,\qquad \sum_{i=0}^m n_i = n_Z\, .
$$
We will employ perturbation theory with small parameter $g_{YM}$ and compute two point functions
to zeroth plus first order. Schematically, there are two types of contributions (corresponding to 
$\alpha$ and $\beta$)
$$
  \langle {\cal O}_1(x_1){\cal O}_2(x_2)\rangle
          = {\alpha +\beta g_{YM}^2\over |x_1-x_2|^{\Delta_0 +g_{YM}^2\Delta_1}}
          = {\alpha +\beta g_{YM}^2\over |x_1-x_2|^{\Delta_0}}\left( 1-g_{YM}^2\Delta_1\log |x_1-x_2|\right)\, .
$$
To obtain $\Delta_1$ we should pick up the $O(g_{YM}^2)$ contribution that 
also comes multiplied by $\log |x_1 - x_2|$. 

At leading order in the $g_{YM}$ expansion the $sl(2)$ sector can be understood as a reduction of 
${\cal N}=4$ super Yang-Mills theory to a sector built using lightcone derivatives of a single complex scalar 
$Z$. The basic two point function we use is
\begin{equation}
  \langle Z^i_j(x)(Z^\dagger)^k_l (y)\rangle = \frac{\delta^i_l\delta^k_j}{|x-y|^2}\, .
\end{equation}
Choose $x^\pm = x^1\pm x^0$. We will consider ${\cal N}=4$ super Yang-Mills theory on the one point
compactification of $R^4$ and will take our two operators to lie at opposite poles of the (conformally
equivalent) $S^4$. The coordinate patch around the south pole has coordinate $x$ while the coordinate
patch around the north pole has coordinate $x'$. We approach the south pole by setting $x^2=x^3=0$ and
taking $x^-=x^+\to 0$ and approach the north pole by setting $x^{\prime 2}=x^{\prime 3}=0$ and
taking $x^{\prime +}=x^{\prime -}\to 0$. The correlators we consider have the form
\begin{equation}
\langle {\cal O}_1 (P){\cal O}_2 (Q)\rangle \equiv \lim_{x,x'\to 0}  \langle {\cal O}_1 (x){\cal O}_2 (x')\rangle\, .
\label{ZamMet}
\end{equation}
We are using a standard trick in conformal field theory: the spacetime dependent two point functions define 
a (spacetime independent) metric on the space of local operators, the Zamalodchikov metric. Knowing the metric 
for arbitrary derivatives of the operators allows us to reconstruct the full spacetime dependence. Working 
with the Zamolodchikov metric will be particularly useful for the problem we consider here. 
See \cite{ginsparg,polchinski,withtom,bhr2} for useful related material.

Under 
\bea
  y^+={1\over x^{\prime -}},\qquad y^-={1\over x^{\prime +}} 
\label{finalchange}
\eea
we have
\bea
  Z(x')\to x^+ x^- Z(x)\, ,
\eea
\bea
{\partial \over\partial x^{\prime -}}Z(x')\to{\partial x^+ \over\partial x^{\prime -}}
{\partial \over\partial x^{+}} \left(  x^+ x^- Z(x) \right) 
\eea
and other obvious generalizations of these formulas for higher derivatives. These follow because $Z$ is a 
conformal primary with dimension 1. Thus, for example
\bea 
\langle Z(P) Z^\dagger (Q)\rangle &&=\lim_{x^\pm,x^{\prime\pm}\to 0}\langle Z^a_b(x) (Z^\dagger)^c_d (x')\rangle\nonumber\\
&&=\lim_{x^\pm\to 0}\lim_{y^\pm\to\infty} y^+ y^- \langle Z(x)Z^\dagger (y)\rangle\nonumber\\
&&=\lim_{x^\pm\to 0}\lim_{y^\pm\to\infty} y^+ y^- {1 \over (x^+-y^+ )(x^- -y^-)}\nonumber\\
&&=1\, . \nonumber
\eea 
Proceeding in this way it is straight forward to argue that
\bea
\langle {1\over n_1 !}\partial^{n_1}_+Z^i_j(P){1\over n_2!}\partial^{n_2}_-{(Z^\dagger)}^k_l(Q)\rangle = \delta^{n_1\, n_2}\delta^i_l\delta^k_j\, ,
\eea
and
$$
  \langle {1\over p_1 !}\partial^{p_1}_+ Z^{i_1}_{j_1}\cdots {1\over p_n!}\partial^{p_n}_+ Z^{i_n}_{j_n}(P) 
          {1\over q_1 !}\partial^{q_1}_-  (Z^\dagger)^{k_1}_{l_1}\cdots 
          {1\over q_n !}\partial^{q_n}_-  (Z^\dagger)^{k_n}_{l_n}(Q)\rangle =\sum_{\sigma\in S_n}\sigma^{I}_{L}(\sigma^{-1})^{K}_{J}
          \prod_{i=1}^n\delta^{p_i q_{\sigma(i)}} \, .
$$
This is a nice formula because it tells us that we can treat $\partial^{p}_+ Z$ as a new type of matrix for each $p$. Thus, we will have
the usual restricted Schur polynomials with one Young diagram for each value of $p$. We will not review restricted Schur polynomials
here. The reader wanting more details may find it useful to consult \cite{de Mello Koch:2007uu,Bhattacharyya:2008rb}.

We study the limit that $n_z,m\sim N$ and $n_Z\gg m$. It is within this subsector of the $sl(2)$ sector that the action of the dilatation
operator simplifies considerably. Our restricted Schur polynomials will be labeled by a Young diagram $R$ with
$n_Z$ boxes and less than $m$ Young diagrams $\{ r_i\} $, $i=0,1,2,...,m$ which each have $n_i$ boxes. The restricted Schur polynomial is
$$
\chi_{R,\{ r_i\}\alpha\beta}(Z^{(0)},Z^{(1)},...,Z^{(m)})=\prod_{k=0}^m {1\over n_k !}\sum_{\sigma\in S_{n_Z}}\chi_{R,\{ r_i\}\alpha\beta}(\sigma )
{\rm Tr}(\sigma \prod_{j=0}^m (Z^{(j)})^{\otimes n_j})\, .
$$
The label $\{ r_i\}\alpha\beta$ specifies an irreducible representation of $S_{n_0}\times S_{n_1}\times\cdots\times S_{n_m}$. It consists of
less than $m$ Young diagrams ($\{ r_i\}$) together with a pair of multiplicity labels ($\alpha\beta$). 
A given $S_{n_0}\times S_{n_1}\times\cdots\times S_{n_m}$
irreducible representation may be subduced more than once; the multiplicity labels tell us which of the degenerate copies are being
used by the restricted character $\chi_{R,\{ r_i\}}(\sigma )$. The free two point function is
\begin{equation}
\langle \chi_{R,\{ r_i\}\alpha\beta}(P)\chi_{S,\{ s_j\}\delta\gamma}^\dagger (Q)\rangle=\delta_{RS}
\delta_{\{ r_i\}\{ s_j\}} \delta_{\alpha\gamma}\delta_{\beta\delta}{{\rm (hooks)}_R\over {\rm (hooks)}_{\{ r_i\} }}f_R\, .
\label{result}
\end{equation}
The delta function $\delta_{\{ r_i\}\{ s_j\}}$ is 1 if the two $S_{n_0}\times S_{n_1}\times\cdots\times S_{n_m}$
irreducible representations specified by $\{ r_i\}$ and $\{ s_j\}$ are identical; multiplicity labels must also 
match - see \cite{Bhattacharyya:2008rb} for more details. The number ${\rm (hooks)}_{R}$ is the product of the hook lengths for Young diagram 
$R$. The number ${\rm (hooks)}_{\{ r_i\} }$ is the product of the ${\rm (hooks)}_{r_i}$, one factor for each of the $r_i$
appearing in $\{ r_i\}$.

\section{Dilatation Operator in $sl(2)$ Subsector}

The one loop contribution to the dilatation operator, in the $sl(2)$ sector, is \cite{beisert}
\bea
\nonumber
  D_1 &&=g_{YM}^2 C^{cd}_{ab}:{\rm Tr}\left(\big[Z^{(a)},Z^{\dagger}_{(c)}\big]\big[Z^{(b)},Z^{\dagger}_{(d)}\big]\right):\\
 &&= g_{YM}^2(C^{cd}_{ab}+C^{dc}_{ab}):{\rm Tr}(Z^{(a)}Z_{(c)}^\dagger Z^{(b)}Z_{(d)}^\dagger):\nonumber\\
 &&-g_{YM}^2(C^{cd}_{ab}+C^{dc}_{ba}):{\rm Tr}(Z^{(a)}Z_{(c)}^\dagger Z_{(d)}^\dagger Z^{(b)}):
\label{dilformula}
\eea
where repeated indices are summed. By cyclicity of the trace we can cycle the two commutators around so 
that our coefficients obey
$$
  C^{cd}_{ab}=C^{dc}_{ba}\, .
$$
When acting on a restricted Schur polynomial, the daggered fields in $D_1$ are to be Wick contracted with fields
in the restricted Schur polynomial so that they act as derivatives
$$
 \left(Z^{\dagger}_{(c)}\right)^i_j \leftrightarrow {d\over \left( dZ^{(c)}\right)^j_i}\, .
$$
Consequently, the one loop dilatation operator acts by removing two fields from the restricted Schur
polynomial and then it puts two others back into the polynomial. Introduce the notation
$$
  c^+ =\sum_{i=1}^{c-1}n_i+1
$$
This is the number of the slot in which the first $Z^{(c)}$ appears. A rather straight forward computation
gives (see \cite{VinceKate,dgm} for related computations with more details)
{\small
\bea
D_1&&\!\!\!\!\!\!\!\!\!\!\chi_{R,\{ r_i\} }(Z,Z^{(1)},...,Z^{(m)})=
g_{YM}^2 (C^{cd}_{ab}+C^{dc}_{ab})
{\prod_k \tilde n_k!\over \prod_l n_l!}\sum_{S,\{ s_i \}}{n_c n_d \over \prod_q d_{s_q}}
\quad \times
\nonumber\\
&&\qquad {\rm Tr}( \sigma (c^+,d^+)P_{R\to \{r_i\}}\sigma^{-1}P_{S\to \{ s_i\}})\chi_{S,\{ s_i\} }(Z,Z^{(1)},...,Z^{(m)})
\nonumber\\
&&-g_{YM}^2 (C^{cd}_{ab}+C^{dc}_{ba})
 {\prod_k\tilde n_k\over \prod_l n_l!}\sum_{R'}\sum_{S,\{ s_i\}}{c_{RR'}d_S n_c n_d\over d_{S'}\prod_q d_{s_q} n_Z}\times
\label{NewDil}\\
&&\times {\rm Tr}(\sigma (n^+,l^+)P_{R\to\{ r_i\}}\sigma^{-1}I_{R'S'}\rho P_{S\to \{ s_i\}}\rho^{-1}(n^+,l^+)I_{S'R'})
 \chi_{S,\{s_i\}}(Z,Z^{(1)},...,Z^{(m)})\, .
\nonumber
\eea
}
The $\tilde n_k$ are the number of impurities $Z^{(k)}$ appearing in $\chi_{S,\{s_i\}}(Z,Z^{(1)},...,Z^{(m)})$.
$\sigma$ in the first term ensures that the fields $Z^{(a)}$ and $Z^{(b)}$ that are inserted are inserted into the correct 
slots. Notice that the first term above does not involve a sum over Young diagrams $R'$ that can be obtained by removing a 
single box from $R$. The structure of this term is new and rather different to the terms obtained when acting with the
dilatation operator in the $su(2)$ sector. In contrast to this, the second term is very similar to terms that arise in
the $su(2)$ sector. In the second term $\sigma$ ensures that the box that is removed corresponds to a $Z$ slot, while $\rho$ 
ensures that the fields $Z^{(a)}$ and $Z^{(b)}$ that are inserted are inserted into the correct slots. $c_{RR'}$ is the factor
of the box that must be removed from $R$ to obtain $R'$. $P_{R\to \{r_i\}}$ is a projection operator projecting from
$R$ to $\{r_i\}$, while $I_{S'R'}$ are intertwiners discussed in detail in \cite{VinceKate,dgm}.  
Equation (\ref{NewDil}) is a new result. 

Recall that we study the limit in which $n_z,m\sim N$ and $n_Z\gg m$. The dilatation operator simplifies considerably
in this limit: since there are a lot more $Z$s than $Z^{(q)}$s with $q>0$ the only term that we need to consider is obtained when 
one of $c,d$ in (\ref{dilformula}) is 0 and one of $a,b$ is 0. In this case (there is a sum on $q$), at leading order in large $N$,
$$
  D_1=2g_{YM}^2(C^{0q}_{0q}+C^{0q}_{q0})\left(:ZZ^\dagger Z^{(q)}Z^\dagger_{(q)}:+:Z^{(q)}Z^\dagger Z Z^\dagger_{(q)}:\right)
$$
$$
   -g_{YM}^2(C^{0q}_{0q}+C^{q0}_{q0})\left(:ZZ^\dagger Z^\dagger_{(q)}Z^{(q)}:+:Z^{(q)}Z^\dagger_{(q)}Z^\dagger Z:\right)
$$
$$
   -g_{YM}^2(C^{0q}_{q0}+C^{q0}_{0q})\left(:Z^{(q)}Z^\dagger Z_{(q)}^\dagger Z:+:ZZ^\dagger_{(q)}Z^\dagger Z^{(q)}:\right)\, .
$$
Given this structure it is clear that processes that change the impurity type are subleading. It is then simplest to assume 
that there is only one type of impurity, $Z^{(q)}$. In this case the label $\{ r_i\}$ contains two Young diagrams.
Using the results of \cite{beisert} we have
$$
  C^{0q}_{0q}=-{1\over 2}h(q)=-{1\over 2}\sum_{i=1}^q {1\over i}
$$
and
$$
  C^{q0}_{0q}+C^{0q}_{q0}={1\over q}\, .
$$

A little work now shows that
$$
  D_1 \chi_{R,(r,s)\alpha\beta }(Z,Z^{(q)})=\sum_{S,(t,u)\gamma\delta}M_{R,(r,s),\alpha\beta\, S,(t,u)\delta\gamma}\chi_{S,(t,u)\delta\gamma}(Z,Z^{(q)})
$$
where
$$
M_{R,(r,s),\alpha\beta\,\, S,(t,u)\delta\gamma} = 2g_{YM}^2 \left( {1\over q}-\sum_{i=1}^q {1\over i}\right)
{nm\over d_r d_s}\chi_{R,(r,s)\beta\gamma}\left( (1,m+1)\right)\delta_{RS}\delta_{(r,s)\, (t,u)}\delta_{\alpha\delta}
$$
$$
- {1\over q}\sum_{R'}{g_{YM}^2 c_{RR'}d_S n m \over d_t d_u (n+m) d_{R'}}
\left[ {\rm Tr}(I_{S'R'} (1,m+1)P_{R\to (r,s)\alpha\beta}I_{R'S'}(1,m+1)P_{S\to (t,u)\gamma\delta})\right.
$$
$$
\left. + {\rm Tr}(I_{S'R'}P_{R\to (r,s)\alpha\beta}(1,m+1)I_{R'S'}P_{S\to (t,u)\gamma\delta}(1,m+1)) \right]
$$
$$
+\sum_{i=1}^q {1\over i}\sum_{R'}{g_{YM}^2 c_{RR'}d_S n m \over d_t d_u (n+m) d_{R'}}
\left[ {\rm Tr}(I_{S'R'} P_{R\to (r,s)\alpha\beta}(1,m+1)I_{R'S'}(1,m+1)P_{S\to (t,u)\gamma\delta})\right.
$$
$$
\left. + {\rm Tr}(I_{S'R'}(1,m+1) P_{R\to (r,s)\alpha\beta}I_{R'S'}P_{S\to (t,u)\gamma\delta}(1,m+1)) \right]\, .
$$
Note the change in notation: we have used $n$ for the number of $Z$s and $m$ for the number of $Z^{(q)}$. Thus,
there are $mq$ covariant derivatives acting. This new notation is far more natural - it fixes the number of impurity
slots to be $m$.

\subsection{The Special Case $q=1$} 

In this case we find
$$
M_{R,(r,s)\alpha\beta\,\, S,(t,u)\delta\gamma} = -\sum_{R'}{g_{YM}^2 c_{RR'}d_S n m \over d_t d_u (n+m) d_{R'}}
\left[ {\rm Tr}(I_{S'R'} (1,m+1)P_{R\to (r,s)\alpha\beta}I_{R'S'}(1,m+1)P_{S\to (t,u)\gamma\delta})\right.
$$
$$
\left. + {\rm Tr}(I_{S'R'}P_{R\to (r,s)\alpha\beta}(1,m+1)I_{R'S'}P_{S\to (t,u)\gamma\delta}(1,m+1)) \right]
$$
$$
+ \sum_{R'}{g_{YM}^2 c_{RR'}d_S n m \over d_t d_u (n+m) d_{R'}}
\left[ {\rm Tr}(I_{S'R'} P_{R\to (r,s)\alpha\beta}(1,m+1)I_{R'S'}(1,m+1)P_{S\to (t,u)\gamma\delta})\right.
$$
$$
\left. + {\rm Tr}(I_{S'R'}(1,m+1) P_{R\to (r,s)\alpha\beta}I_{R'S'}P_{S\to (t,u)\gamma\delta}(1,m+1)) \right]
$$
$$
= -\sum_{R'}{g_{YM}^2 c_{RR'}d_S n m \over d_t d_u (n+m) d_{R'}}
{\rm Tr}(I_{S'R'} \big[(1,m+1),P_{R\to (r,s)\alpha\beta}\big]I_{R'S'}\big[(1,m+1),P_{S\to (t,u)\gamma\delta}\big])\, .
$$
This is identical to the action of the dilatation operator in the $su(2)$ sector (see formula (2.3) of \cite{VinceKate})!
Denote this as $M_{R,(r,s)\alpha\beta\,\, S,(t,u)\delta\gamma}^{\rm su(2)}$.

\subsection{The General Case of $q>1$} 

In this case we can write
$$
M_{R,(r,s)\alpha\beta\,\, S,(t,u)\delta\gamma} ={1\over q} M_{R,(r,s)\alpha\beta\,\, S,(t,u)\delta\gamma}^{\rm su(2)} +
\delta M_{R,(r,s)\alpha\beta\,\, S,(t,u)\delta\gamma}
$$
where
$$
 \delta M_{R,(r,s)\alpha\beta\,\, S,(t,u)\delta\gamma}=2g_{YM}^2\left({1\over q}-\sum_{i=1}^q {1\over i}\right)\delta_{RS}
\delta_{(r,s)(t,u)}\delta_{\alpha\delta}{nm\over d_r d_s}\chi_{R,(r,s)\beta\gamma}\left((1,m+1)\right)
$$
$$
-g_{YM}^2\left({1\over q}-\sum_{i=1}^q {1\over i}\right)\sum_{R'}{c_{RR'}d_S n m \over d_t d_u (n+m) d_{R'}}
\left[
{\rm Tr}(I_{S'R'} P_{R\to (r,s)\alpha\beta}(1,m+1)I_{R'S'}(1,m+1)P_{S\to (t,u)\gamma\delta})\right.
$$
$$ + \left. {\rm Tr}(I_{S'R'} (1,m+1)P_{R\to (r,s)\alpha\beta}I_{R'S'}P_{S\to (t,u)\gamma\delta}(1,m+1))
\right]\, .
$$
We will sketch how to evaluate $\delta M_{R,(r,s)\alpha\beta\,\, S,(t,u)\delta\gamma}$. First, recall the
action of $E_{ij}$ in the fundamental representation of $u(N)$ on a vector $v_k$, which is a vector of zeros
except for a 1 in the $k^{\rm th}$ position, is
$$
  E_{ij}v_k = \delta_{jk}v_i\, .
$$
The symmetric group element $(1,m+1)$ swaps the vectors in the first and $m+1$th slots. Given the above action
of $E_{ij}$ we can write $(1,m+1)={\rm Tr}(E^{(1)}E^{(m+1)})$. Further, the projector $P_{R\to (r,s)\beta\gamma}$
can be factored into an operator acting on the impurity slots ($p_{R\to(r,s)\beta\gamma}$) and an operator acting 
on the $Z$ slots (${\bf 1}_{r}$)
$$
P_{R\to (r,s)\beta\gamma}=p_{R\to(r,s)\beta\gamma}\otimes {\bf 1}_{r}\, .
$$
The first term in $\delta M$ can now be rewritten using
$$
{nm\over d_r d_s}\chi_{R,(r,s)\beta\gamma}\left((1,m+1)\right)
=\sum_{i\, j} {nm\over d_r d_s}{\rm Tr}(p_{R\to(r,s)\beta\gamma} E^{(1)}_{ij}){\rm Tr}({\bf 1}_{r}E^{(m+1)}_{ji})
$$
$$
=\sum_{i} {nm\over d_r d_s}{\rm Tr}(p_{R\to(r,s)\beta\gamma} E^{(1)}_{ii}){\rm Tr}({\bf 1}_{r}E^{(m+1)}_{ii})\, .
$$

Next, consider the second term in $\delta M$ (to get $R'$ from $R$ remove box in row $j$ and to get $S'$ from 
$S$ remove box in row $i$)
$$
\sum_{R'}{c_{RR'}d_S n m \over d_t d_u (n+m) d_{R'}}
{\rm Tr}(I_{S'R'} P_{R\to (r,s)\alpha\beta}(1,m+1)I_{R'S'}(1,m+1)P_{S\to (t,u)\gamma\delta})
$$
$$
= \sum_{i}{c_{RR'}d_S n m \over d_t d_u (n+m) d_{R'}}
{\rm Tr}(E_{ij}^{(1)} P_{R\to (r,s)\alpha\beta}E_{ji}^{(m+1)}P_{S\to (t,u)\gamma\delta})
$$
$$
= \sum_{i}{c_{RR'}d_S n m \over d_t d_u (n+m) d_{R'}}
{\rm Tr}(E_{ij}^{(1)} p_{R\to (r,s)\alpha\beta}p_{S\to (t,u)\gamma\delta})
 {\rm Tr}({\bf 1}_{r}E_{ji}^{(m+1)}{\bf 1}_{t})
$$
$$
= \delta_{RS}\delta_{(r,s)\, (t,u)}\delta_{\beta\gamma}\sum_{i}{c_{RR'}d_R n m \over d_t d_u (n+m) d_{R'}}
{\rm Tr}(E_{ii}^{(1)} p_{R\to (r,s)\alpha\delta})
 {\rm Tr}({\bf 1}_{r}E_{ii}^{(m+1)})\, .
$$
To proceed note that
$$
  c_{RR'}=N+{{\rm hooks}_R\over {\rm hooks}_{R'}}
$$
and treat these two terms separately. The term that comes from the $N$ gives us
$$
\delta_{RS}\delta_{(r,s)\, (t,u)}\delta_{\beta\gamma}\sum_{i}{Nd_R n m \over d_r d_s (n+m) d_{R'}}
{\rm Tr}(E_{ii}^{(1)} p_{R\to (r,s)\alpha\delta})
 {\rm Tr}({\bf 1}_{r}E_{ii}^{(m+1)})
$$
$$
=\delta_{RS}\delta_{(r,s)\, (t,u)}\delta_{\beta\gamma}\sum_{i}{N m \over d_s }
{\rm Tr}(E_{ii}^{(1)} p_{R\to (r,s)\alpha\delta})
$$
$$
=\delta_{RS}\delta_{(r,s)\, (t,u)}\delta_{\beta\gamma}\delta_{\alpha\delta}N m\, .
$$
The remaining piece is
$$
\delta_{RS}\delta_{(r,s)\, (t,u)}\delta_{\beta\gamma}\sum_{i}{d_R n m \over d_r d_s (n+m) d_{R'}}{{\rm hooks}_R\over{\rm hooks}_{R'}}
{\rm Tr}(E_{ii}^{(1)} p_{R\to (r,s)\alpha\delta})
 {\rm Tr}({\bf 1}_{r}E_{ii}^{(m+1)})
$$
$$
=\delta_{RS}\delta_{(r,s)\, (t,u)}\delta_{\beta\gamma}\sum_{i}{ n m \over d_r d_s }
{\rm Tr}(E_{ii}^{(1)} p_{R\to (r,s)\alpha\delta})
 {\rm Tr}({\bf 1}_{r}E_{ii}^{(m+1)})\, .
$$
The third term can be dealt with in a very similar way. Together these results are enough to evaluate 
$$
 \delta M_{R,(r,s)\alpha\beta\,\, S,(t,u)\delta\gamma}=2\delta_{RS}\delta_{(r,s)\, (t,u)}\delta_{\beta\gamma}\delta_{\alpha\delta}
                                                        \left(\sum_{i=1}^q {1\over i}-{1\over q}\right)g_{YM}^2 Nm\, .
$$
This just adds a constant times the identity to the $su(2)$ dilatation operator so that we find exactly the same 
eigenvalue problem as for the $su(2)$ sector except that, for $q>1$, 
there is a shift due to the presence of $\delta M_{R,(r,s)\alpha\beta\,\, S,(t,u)\delta\gamma}$. 
This shift is positive as it must be: a negative shift would have produced operators with a dimension less than their
${\cal R}$-charge which is not possible in a unitary conformal field theory.

\section{Discussion}

We have computed the action of the dilatation operator on restricted Schur polynomials that are dual to systems of excited
AdS giant gravitons. The open string excitations have an angular momentum along an $S^1\subset$AdS$_5$. For a similar project
which considers a single AdS giant see \cite{correa}. Our study has shown that diagonalizing the action of the dilatation operator
in the $sl(2)$ sector can be reduced to the problem of diagonalizing the action of the dilatation operator in the $su(2)$ sector.
For the $su(2)$ sector problem it is known that the Gauss Law emerges at one loop and the spectrum reduces to that of a set
of decoupled oscillators. Our results show that in the $sl(2)$ sector we again find that the Gauss Law emerges and the
spectrum is again given by a set of decoupled oscillators. Our results extend and complement those of \cite{Koch:2010gp,VinceKate,bhw,dgm}. 
Taken together, these results are very suggestive that there are indeed large $N$ but non-planar limits in which the 
anomalous dimensions continue to be governed by an integrable system. It would be fascinating to study this possibility further.

There is a constant shift for the anomalous dimensions of the operators (as compared to the $su(2)$ result) if $q> 1$. 
This is quite natural: the condition to be a BPS state is that we have the ``biggest possible {\cal R}-charge'' for a given 
dimension $\Delta$. To get the largest {\cal R}-charge, one needs to symmetrize on all indices. For $q>1$ we have a clump of 
$q$ covariant derivatives i.e. $q$ indices are stuck together. Thus for $q>1$ it is not possible to completely symmetrize on 
all indices, so that there are no BPS operators.

We have focused on a single species of impurity. However, using the results of \cite{steph} we know it is not difficult to add 
more impurity flavors. This uses the fact that at leading order in large $N$ there are no impurity number changing processes.

Although we have focused on restricted Schur polynomials in this article, they are not the only basis for local gauge invariant 
operators of a matrix model. Another interesting basis to consider is the Brauer basis\cite{Kimura:2007wy,Kimura:2009wy}. There 
is an elegant construction of a class of BPS operators \cite{yusuke} in which the natural $N$ dependence appearing in the definition of the 
operator\cite{heslop} is reproduced by the Brauer algebra projectors\cite{yusuke}. Another natural approach 
is to adopt a basis that has sharp quantum numbers for the global symmetries of the theory\cite{Brown:2007xh,bhr2}. The action of 
the anomalous dimension operator in this sharp quantum number basis is very similar to the action in the restricted Schur 
basis\cite{Brown:2008rs}. Finally, for a rather general approach which correctly counts and constructs the weak coupling BPS 
operators see\cite{jurgis}. Are there signals of non-planar integrability in these more general bases?

\noindent
{\it Acknowledgements:}
This work is based upon research supported by the South African Research Chairs
Initiative of the Department of Science and Technology and National Research Foundation.
Any opinion, findings and conclusions or recommendations expressed in this material
are those of the authors and therefore the NRF and DST do not accept any liability
with regard thereto.

\end{document}